# The Last Scientific Revolution

Andrei P. KIRILYUK[*]

Solid State Theory Department, Institute of Metal Physics
36 Vernadsky Avenue, 03142 Kiev-142, Ukraine

Critically growing problems of fundamental science organisation and content are analysed with examples from physics and emerging interdisciplinary fields. Their origin is specified and new science structure (organisation and content) is proposed as a unified solution.

## 1. The End of Lie, or What's Wrong With Science

Whereas today's spectacular technologic progress seems to strongly confirm the utility of underlying scientific activities, the modern state of fundamental science itself shows catastrophically accumulating degradation signs, including both knowledge content and organisation/practice [1-49]. That striking contradiction implies that we are close to a deeply rooted *change* in the whole system of human knowledge directly involving its *fundamental nature* and application *quality* rather than only superficial, practically based influences of empirical technology, social tendencies, etc. Science problems, in their modern form, have started appearing in the 20$^{th}$ century, together with accelerated science development itself [50-55], but their current culmination and now already long-lasting, well-defined crisis clearly designate the advent of the biggest ever *scientific revolution* involving not only serious changes in special knowledge content but also its *qualitatively new character*, meaning and role [33-39,44-49,56]. We specify below that situation, including today's science problems, related development issues and objectively substantiated propositions for sustainable progress.

Summarising *critically growing problems of modern science*, it would be not out of place to begin with the *internal science estimate* by its practitioners and dynamics of its "human dimensions". Increasingly dominating *mediocrity* of results, human choices and relations in today's *fundamental science* is expressed in a huge variety of public or private opinions (e.g. [1-39,42-46]) and presents a striking contrast not only to simultaneous triumph of empirically advancing technologies, but also to a previous, very recent (decades-old) and visibly huge, euphoric success of the same science, with ever brighter perspectives appealingly looming ahead. But more than ever "pride goes before destruction", and the haughty spirit of seemingly omnipotent knowledge is "suddenly" transformed now into a dirty fight of vain ambitions accompanied by the "discovery" of omnipresent ethical decay in science practice (direct fraud or "officially permitted" lie, organisational corruption) [18-38] rather than any novel truth about reality. We reveal below the exact, rigorously specified reason for, and *true* meaning of, such dramatic "end of science" (cf. [3]), as well as much more positive perspectives of knowledge development far beyond its now ending, *unitary* level [35].

We show that scandalously stagnating old scientific problems, its supernatural "mysteries", and increasingly accumulating, catastrophically big new paradoxes result from the *artificially limited* scheme of officially dominating, "positivistic" science approach, where the unreduced, interaction-driven, *dynamically multivalued* reality is replaced by its artificially fixed, *effectively zero-dimensional (point-like) projection* [35-37]. That huge, *maximum possible* simplification of reality within the *subjectively imposed* unitary doctrine explains both its visible (though always strongly *incomplete*) success at the *lowest* levels of *complex* (multivalued) world dynamics and even more evident (and this time *complete*) failure to provide objective world description at *higher complexity levels*, marking the border between "exact" and "natural" sciences (let alone "humanities" and arts) which is only formally postulated in the official knowledge framework. Thus *scientifically* specified origin of conventional science limitations exceeds essentially their existing empirical descriptions (e.g. [3-6,19-21,39,50,55]) that often correctly present various aspects of the resulting knowledge degradation, but fail to trace its genuine meaning and related perspectives of positive change.

---
[*] Address for correspondence: A.P. Kirilyuk, Post Box 115, 01030 Kiev-30, Ukraine.
E-mail: Andrei.Kirilyuk@Gmail.com.

It is not surprising that a completely deformed, over-simplified picture of reality, not even attempting at its genuine, causally complete understanding, evolves into equally strongly simplified, *authoritarian organisation* and *criminal practice* of science today (see e.g. [7-14,22-34]). Indeed, what is the sense to ask for more ethics in research (i.e. in the whole self-protective enterprise of today's bankrupt scholar science), if this *particular* kind of knowledge is *based*, starting from its most fundamental levels, on explicit and evident trickery of supernatural "mysteries", *officially postulated* in an allegedly "objective" and "rigorous" doctrine? The effectively zero-dimensional, *zero-complexity* "model" of dynamically multivalued, *high-complexity* reality, totally dominating and *artificially imposed* in all research and teaching institutions, is none other than officially legalised, *maximum* possible *lie* about the real world structure and dynamics. All previous and modern *science wars* and *unethical* cases of science practice are but manifestations of that single, basic *departure from the truth*, the latter always remaining, however, the officially announced purpose of science. We show how the necessary transition to the new, unreduced form of knowledge – exemplified by the *already created* framework of the *universal science of complexity* [35-37,57-66] – naturally involves the related deep change of science organisation and practice towards an *explicitly creative* system of knowledge production and dissemination.

## 2. The End of Unitary Thinking

### 2.1. Knowledge without Explanation: Postulated Blunders of Official Science

#### 2.1.1. Mathematics of reality vs. mathematical imitations of reality

As noted above, organisational, social and other "human" problems of modern science are directly related to the unitary, strongly and explicitly reduced framework of its content, and as the latter is determined by the mathematical basis of science, one cannot avoid analysing its limitations that underlie and illustrate all higher-level features of knowledge. The *deliberate rejection* of a search for *realistic, complete explanation* of natural phenomena in terms of *natural* entities in favour of *purely abstract* "model" adjustment to *quantitative* results of *selective* measurements is the real, now totally dominating basis of official science doctrine, also known as "positivistic science" (due to explicit emphasis of the concept by Auguste Compte) and stemming from Isaac Newton's approach and attitude ("hypotheses non fingo"). It is that *very special*, strongly and *artificially* limited "paradigm" that was imposed on the *whole* body of knowledge since the apparent "great" success of Newton's model, which explains also its modern "end" accompanied by a catastrophically growing multitude of difficult, practically "unsolvable" problems and related "distrust of science". *Another* kind of science oriented to the *causally complete* understanding of reality was founded by René Descartes half a century before Newton, but was later practically totally excluded from official science practice despite the accumulating difficulties of Newtonian "science without explanation". That "terrible mistake" of conventional knowledge involves transformation of mathematical *tools* (or *language*) of science into its self-important, absolute and therefore dominating *purpose*, so that the whole tangible reality is finally replaced in that "positivistic" doctrine by purely abstract world of mathematical, imitative and immaterial structures (see e.g. [67-75]). But is it a problem of mathematics as such and can the unreduced, tangible reality be *exactly* reproduced by a mathematically rigorous language of science? As shown in detail in the universal science of complexity [35-37,57-66], the answer is positive and it is due to essential and well-specified *extension* of the basically *wrong*, artificially restricted use of mathematics in conventional science.

Indeed, if we apply the same mathematical tools in their unreduced, truly correct version to the same natural phenomena, we immediately discover that the difficulties of usual science description are due to the following *major blunders* of conventional mathematics application arising from a *subjective* and artificially imposed bias towards maximum, unjustified, mechanistic *simplicity*.

# The Last Scientific Revolution

(1) **Uniqueness theorems**. Statements of standard mathematical theorems about *uniqueness* of generic problem solution, serving as a "well-established" basis for the whole scholar mathematics and its applications, are plainly and totally *erroneous*, just for generic, at least basically realistic cases of interaction problem (and for *all real* interactions). Indeed, as a rigorous problem analysis and its physically transparent interpretation convincingly show [35-37,57-66], every unreduced interaction process, underlying any observed phenomenon or system dynamics, gives rise to *dynamic multivaluedness*, or *redundance*, of problem solutions, where the problem actually has many locally "complete" and therefore *mutually incompatible* solutions called (system) *realisations* that are forced, therefore, by the driving interaction itself, to *permanently replace each other* in a *causally* (and *truly*) *random* (alias *chaotic*) order thus defined. Only unrealistically simplified, strictly one-dimensional (and timeless) interaction problem can have a unique solution. Usual mathematics incorrectly extends this situation to real cases by neglecting omnipresent real-problem instabilities with respect to multiple possible scattering ways implying multivaluedness of the effective interaction potential, as opposed to single-valued potential assumption (often tacitly) imposed in conventional theorems. Deeply related to this feature is a totally subjective inclination of usual mathematics for a basically *smooth* (unitary) and *closed* ("exact" or "separable"), i.e. again *simple*, form of all its constructions, in striking contrast to the observed configuration and dynamics of natural systems.[†] In fact, usual, *dynamically single-valued* solution of a *real* problem does *not* even *exist* as such.

(2) **Self-identity postulate**. Irrespective of detailed statements and structures, a single, omnipresent, implicit postulate underlies all mathematical constructions of usual science, the "evident" assumption about self-identity, $\mathfrak{A} = \mathfrak{A}$, of any given structure $\mathfrak{A}$ (including, of course, its possible time dependence, $\mathfrak{A}(t) = \mathfrak{A}(t)$). The above dynamic multivaluedness and related permanent, unceasing realisation change of *any* system structure shows that real world objects and their *correct* mathematical images are *not* self-identical, $\mathfrak{A} \neq \mathfrak{A}$. Moreover, the same property of unreduced, multivalued system dynamics gives rise to the *rigorously*, universally defined notions of *event* (realisation change), *emergence* (new realisation formation) and *time*, $\mathfrak{A} \neq \mathfrak{A} \Rightarrow \mathfrak{A}(t)$, the latter acquiring now its realistic, uneven, unstoppable and causally irreversible origin [35-37,58,60-62,64].

(3) **Absence of material quality**. Again irrespective of details, there is no possibility in usual mathematics applications to express the *tangible material quality*, or "texture", of a real object as a whole, in its unreduced version perceived through experience. Usual mathematical "models" of reality can propose instead only immaterial, over-simplified, "ideal" images supposed to properly *imitate* real objects whose particular measured characteristics (such as density, roughness, etc.) can be expressed by respective quantities where necessary. However, one can never obtain anything closely resembling a real object as a whole, with its detailed structure and dynamics. In fact, usual mathematics is missing completely even such problem formulation. The causally complete framework of the universal science of complexity provides that missing *rigorous* description of material quality in the form of *dynamically multivalued* (and thus chaotically changing) *entanglement* of interacting system components organised, in addition, in a hierarchy of levels of *dynamically probabilistic fractal* [35,36,64,65] representing the *real* object structure.[‡] Moreover, the same framework shows *why* one *cannot* obtain a correct quality expression in usual mathematics in principle: this is because its invariable simplification of real interaction process by artificial cutting of emerging dynamical links (in the search for an "exact", closed-form solution within a version of "perturbation theory") just automatically kills both key features of the unreduced solution, dynamic multivaluedness and fractal entanglement. It is interesting that in contrast to apparently desirable property of realism, official mathematics applications in modern science (dominated by so-called "mathematical physics") seem to be *proud* of their immaterial nature *deliberately* favoured due to a very special assumption about the fundamental basis of the observed world structure [67-75], which reveals a strangely subjective, doctrinaire and almost "religious" attitude behind the allegedly "objective" form of knowledge officially supported and absolutely dominating in all "secular" educational and scientific institutions.

---

[†] One should emphasize the key difference of our *dynamic* multivaluedness from usual, mechanistic multivaluedness (also referred to as "multistability") where "multivalued" solution components represent but an internal structure of a *single*, dynamically smooth (unitary) system realisation. Those components are *regularly* taken by the system along its *formally* advancing trajectory (or state evolution) in an *abstract*, rather than real, space (such as "phase space" in Hamiltonian dynamics). All the "multiple", *coexisting* "attractors" of scholar "nonlinear science" describe but usual, basically *linear* trajectory configurations in an abstract space, corresponding to the *unique* problem solution, though maybe having a very "intricate" and therefore "nonintegrable" detailed structure.

[‡] Note that usual fractals do not possess the basic property of dynamic multivaluedness and are not obtained as exact solutions of respective dynamic equations (i.e. as a result of system interaction development). Therefore, they do not show crucially important features of dynamic entanglement of system components and unceasing chaotic change at all structural levels just underlying the correct expression of material quality by the dynamically probabilistic fractal.

(4) ***Absence of genuine randomness***. Dynamically single-valued, "exact", or closed solutions of scholar mathematics do not leave any place for genuine *randomness*, which leads to multiple conceptual and practical difficulties with various related notions and ideas, such as *nonintegrability*, *nonseparability*, *noncomputability*, *chaos*, *uncertainty (indeterminacy)*, *undecidability*, broken symmetry, free will, time flow, etc. (see e.g. [75]). In connection with the above items (1) and (2), the universal science of complexity provides the *unified* solution to all those problems due to a clearly specified origin of *true* randomness in the *dynamically multivalued* structure of the unreduced, *mathematically correct* problem solution [35-37,57-66]. This is important logical closure of the unreduced mathematics of real-world complexity, contrasting with growing "loss of certainty" of usual mathematics [76]: the realistic mathematics of complexity is quite certain and complete, transforming *any* problem into a *solvable* (integrable) one, but the obtained solution *structure* is as intrinsically, dynamically "fuzzy" (irregular) and diverse (changing) as real world phenomena.

(5) ***Continuity or false, mechanistic discreteness***. Continuity, or homogeneity (unitarity) of conventional mathematical structures is a fundamental consequence of their dynamically single-valued, non-dynamic origin. In order to account for the observed uneven patterns, unitary theory introduces artificial, mechanistic discreteness showing the opposite, equally unreal limit of "infinitely sharp" ruptures. The unreduced, dynamically multivalued solution of a real problem reveals instead the phenomenon of *dynamic discreteness* (or *causal quantisation*), where any system dynamics and resulting structures are obtained as continuous but *qualitatively uneven* sequences of system realisations and *transitions* between them occurring through system *reconfiguration* within its special, "intermediate" realisation [35,36,60]. In that way, one gets the causally quantised structure of physically real, tangible *space* and equally real, but immaterial, irreversibly flowing *time*. Many related, major structures and properties of unitary mathematics appear now as only rough (at best), basically wrong approximations to nonunitary structure of reality, including unitarity itself (e.g. of quantum dynamics), continuity *and* (mechanistic) discontinuity, calculus (including its "discrete" and "discontinuous" versions), evolution operators, symmetry operators, any unitary operators, Lyapunov exponents (and related chaoticity imitations), path integrals, statistical theories, etc.

It is important that while the fundamental problems (1)-(5) of conventional mathematics applications look different within its reduced framework, they acquire a *unified origin and solution* within the causally complete mathematical structure of dynamic complexity. That unification has explicit manifestation in the fact that the whole world dynamics, in its unreduced, *totally realistic* version is presented now as a *single*, dynamically unified structure of *dynamically probabilistic fractal*, while all its properties, now *rigorously derived* and causally extending usual, mechanistically imposed laws and postulates, are unified by the *universal symmetry of complexity* [35-37,57,58].§ Correspondingly, the obtained results refer to *all* fields of knowledge, rather than only "exact" or "natural" sciences. Even though various knowledge fields remain *irreducibly separated* in usual science, mathematics is always viewed as a "universal language of science". Universal science of complexity reveals the ultimate completion of that property, where the extended mathematics of complexity does provide a *causally complete* description of *any* system behaviour and ensures coherent transition between different systems and levels of complex world dynamics.

---

§ Universal symmetry of complexity provides a unified solution to another "eternal", unsolvable problem of usual mathematics applications, that of systematic deviations of always "too irregular" real structures from "simple", regularly structured symmetries of unitary theory. The latter then resorts to one of its characteristic "tricks" and introduces the idea of "spontaneously broken symmetry", where one states that a symmetry in question does exist, but if it is not observed in real system structure and dynamics, this is because it is "spontaneously broken" (starting already from only one direction of real time flow, or entropy growth). That unitary "symmetry without symmetry" enters a sad list of other major "contradictions" of unitary doctrine, such as chaos without chaos and complexity without complexity (see also section 2.1.4 below). Universal science of complexity solves this symmetry problem by revealing the *irregular structure* of the real symmetry of nature, so that the symmetry of complexity remains always *exact* but it connects system configurations (realisations) which are *irregularly different* from one another.

## 2.1.2. Purely abstract world: fundamental physics without truth and the end of "unreasonable effectiveness of mathematics"

There is certainly a direct link between the strongly *limited*, evidently incorrect basis of usual mathematics applications in "exact" sciences (see the previous section) and its absolute *domination* in the *purely abstract* world picture used in conventional science. Indeed, all those "complicated details" of real world dynamics, which are boldly rejected (without any sound justification) in the standard science approach, just make the *essential* difference between abstract "models" of the latter and reality they pretend to describe. Starting with Newtonian "hypotheses non fingo", that very special understanding of objective world description of usual, "positivistic" science doctrine finds its apparent confirmation by successful applications to properly selected, simple and "smooth" enough phenomena and structures. Even in those particular cases, however, there is a number of strong, persisting "mysteries" and "difficulties" often directly related to the real origin of entities and properties just deliberately omitted in mechanistic science (e.g. the origin of gravity, mass, equivalence between its gravitational and inertial manifestations, time and space in Newtonian mechanics, canonical "quantum mysteries" and relativity postulates in the "new physics", etc.). There is no surprise that the proportion of "mysteries" grows catastrophically for ever more perfect observation results, where all those rejected "details" become visible and often constitute the "main effect". At that stage, attained in science in the middle of the $20^{th}$ century, the famous "unreasonable effectiveness of mathematics in the physical sciences" [77] (see also [74,75]) turns into even more impressive *in*effectiveness, where the "advanced" unitary mathematics plays its increasingly esoteric games without any relation to real world structure (string theory, loop quantum gravity, etc.). However, it's enough to consider real interaction processes rigorously, without usual illegal "tricks" of perturbation theory, in order to obtain quite realistic, truly exact presentation of reality using formally the *same* mathematical tools (but now in a different, truly *correct way*, see items (1)-(5) in the previous section).

There is no surprise, therefore, that the characteristic "crisis in physics" giving rise to *officially* successful science revolution at the beginning of $20^{th}$ century is strangely reproduced today, after a hundred years of apparently quite prosperous application and development of the "new physics" results. Fundamentally incorrect, purely abstract "mathematical physics" artificially established as a single possible approach in science despite its intrinsic "unsolvable problems" and persisting irrational "mysteries" still cannot provide a truly sustainable science basis. The old doubts and search for a "genuine", truly consistent science foundation reappear again and take now ever more "global" proportions, such as the widely acknowledged "crisis in cosmology" [78,79], intense multiplication of "invisible" entities (hidden space dimensions, dark energy and matter, "theoretically" needed but experimentally absent "supersymmetric" particle partners, etc.), or a deep impasse in the whole fundamental science development known as the "end of science" [3]. And while the leading priests of mathematical physics are still busily disputing the "absolute" advantages of their "personal" abstractions (e.g. [45,68]), it becomes increasingly evident that any, even most "elegant" mathematical trickery cannot solve the problem and the fundamental science development can be restarted and prosperously continue only after a decisive transition to the unreduced, totally realistic vision and description of reality.[**]

---

[**] As persisting problems of usual mathematics applications compromise the "unreasonable effectiveness" thesis in an ever more obvious way, an "additional" argument in their favour is advanced, that of intrinsic, "superior" *beauty* of pure mathematical constructions which *cannot* be useless in principle and therefore *should* find one or another reflexion in real system behaviour (e.g. [74,75]). That ill-defined beauty of *abstract* constructions was even proposed (allegedly by Paul Dirac) as an independent *criterion* of their validity for *real* world description. However, while mathematical physics applications degrade from canonical quantum mysteries to the totally self-absorbed modern branches of string theory or quantum gravity, *their* abstract "beauty" [68] advanced as the *only* remaining criterion of truth and actually appreciated by an *extremely narrow* group of devoted adherents, shows that probably the underlying aesthetical (let alone ethical) standards themselves are not as universal as it is usually assumed. At least one can see a direct contradiction between *that* kind of abstract aesthetical climax of *over-simplified* constructions and undeniable criteria of beauty of "usual", directly perceived reality, where *more complex*, involved and variable objects appear as more beautiful ones (see also [35] for the rigorous substantiation of that relation).

A.P. KirilyukIn the meanwhile, all the extremely costly experiments involving accelerators, satellites and related efforts of industrial scale are always based on those purely abstract concepts that explicitly fail to produce at least a generally consistent picture of reality. One can mention such purely mathematical, even theoretically disputable constructions as Higgs bosons, supersymmetric partners of "usual" particles, various candidates for "dark matter" particles and "dark energy" sources, gravitational waves, black holes, etc. It should be emphasized that those officially accepted (and uniquely supported) schemes show multiple, evident deficiencies already in theory and still they are used as a *single possible* basis for those huge experiments involving hundreds and thousands of highly qualified professionals.†† And when the performed expensive trials of the bankrupt concepts "prosperously" and inevitably fail, one after another (no found expected "superpartners" for ordinary particles, nor gravity modifications due to "hidden dimensions", nor esoteric "dark matter" candidates, etc.), it changes nothing in the accepted practice and theories of scholar science: while personal incomes of the failing enterprise chiefs continue to grow without limits, any reality-based, *causally complete* world description (e.g. [35,59-62]) is excluded from any support at all, despite its clear, though even unintentional, *confirmation* by the same experimental data [60].

However, even huge material losses and impossibility to initiate real problem solution can be not the most serious consequences of such "unlimited" deviation from elementary criteria of truth in science. As the purely empirical, technical science possibilities grow at a spectacular rate, their power exceeds now the *whole* range of natural structure complexity [35,36,60]. Correspondingly, arbitrary application of those empirical tools based on illusive mathematical structures and now multiply disproved postulate of their "unreasonable effectiveness" is practically equivalent to the *premeditated destruction* of those *real* structures, with unpredictable consequences but guaranteed failure of "theory confirmation by experiment". The omnipotent tsars of official science are well aware of the related dangers (see e.g. [39]), but they continue to impose their "old good" trial-and-error method beyond the well-specified limits of its applicability and any reasonable efficiency.

In that way, purely abstract structures can lead indeed to quite tangible, negative consequences for the real world they miss to describe but can effectively destroy. A part of that destruction already clearly appears in physics in the form of practically lost interest of public and related lack of creative young researchers, which only amplifies the crisis of science content and its corrupt organisation practices. It's clearly a time for revolution: what else can reverse those deadly tendencies and transform the current deepening crisis into a sustainable progress?

### 2.1.3. Quantum computers, nano-technology, and other "applied" giga-frauds

Whereas growing difficulties of basic science acquire a fundamental, inevitable origin and consistent explanation (sections 2.1.1-2), it remains to hope that scholar research can have brighter perspectives in its more applied aspects, exemplified by recently appeared "hot" fields of quantum computation, nano-bio-technology, thermonuclear fusion revival, and various "computer science" applications, from new materials design to climate simulations. Closer examination of those billion-worth new "advances" shows, however, that conventional science has quickly degraded from inconsistent imitations of reality to open "intellectual" fraud based on shamelessly "strong" promises that can never be realised, according to undeniable, multiply confirmed laws of the same science.

Thus, unitary *quantum computation* idea, consuming in the last years practically the whole volume of quantum physics and related research (it's enough to have a look at the paper list in quant-ph section of arXiv.org), provides a typical example of that strange combination of *strong* doubts about its practical realisation and ever growing publicity and investments into the extremely dubious enterprise. Indeed, even its active participants openly acknowledge that the most probable expected result of the whole activity is that full-scale quantum computers cannot be built [81]. There are numerous (but "strangely" ignored) particular doubts in fundamental quantum computer feasibility (see [36, section 2] and references therein, [82]). And finally there is a *causally complete* analysis of the universal science of complexity [36] that shows, within a *realistically extended* picture of quantum behaviour (including *genuine* quantum chaos), why exactly quantum computers cannot fulfil their promise even under most "ideal" conditions of their operation. It is easy to see that such causally substantiated conclusion simply confirms (and now realistically *explains*) standard quantum postulates (and other fundamental laws, including entropy growth), which are already multiply confirmed experimentally and *contradict* the very idea of unitary quantum computation [36]. It is the scandalously abusive play on *supernatural* "quantum mysteries" of official "rigorous" science ("multiverse interpretations", etc.) used now for invention of *real*, practically efficient devices that has permitted such incredible (and ever growing) deviation from elementary consistency and honesty. But why can such ultimately perverted activity continue in all the "best" scientific institutions and programmes? It can simply because some officially "leading" scientists have their purely subjective and absolutely unbalanced preference for the underlying manipulation with abstract symbols and "fantastic" promises, while the "embedding" system of science organisation has neither real possibilities, nor interests necessary for critical limitation of such abuses (see section 2.2). Such is another *real* result of the "unreasonable effectiveness of mathematics" (section 2.1.2).

---

†† The failure itself of the dominating "standard model" of particle physics is interpreted as extraordinary possibilities of its development ... within the same model (see e.g. [80])! When they cannot deny any more the total fiasco of their "best possible" doctrine, official science leaders shamelessly transform it into another occasion to ask for new, heavy expenditures for experiments based on the discredited concepts. No limits for such kind of "science", indeed!

Yet much larger modern science "bubble", that of *nanotechnology*, is physically close to quantum computation case, but is actually based on a much less "scientific", mainly publicity-driven trickery. Starting from the evident Feynman's blunder [83] about "plenty of room at the bottom" (directly contradicting major quantum laws), the nanotechnology affair quickly took the scale of unlimited science-fiction hype [84-86] that has received, however, a strangely generous and "top-level" support from all major sources [87]. However, similar to quantum computer case, none of the "fantastic" promises has led to a really novel result or application, despite many years of very intense efforts, "nice pictures" of "small, tricky structures" and continuing multi-billion investment (if only one avoids purely terminological tricks, very popular in this "prosperous" field, when e.g. former computer *micro*-chips are now *classified* as *nano*technological products, just because their details can be as small as a hundred nanometres). It finally becomes evident that the real, practical reason for that bizarre giga-fraud so easily accepted by the most prestigious institutions is the rapid shrinking of the formerly extremely large and prosperous field of solid-state physics (actually due to successful technological applications), whose adherents has found "nanotechnology" as an efficient replacement for their disappearing financial support. The story of that another "science without science" is especially disappointing because the truly scientific, fundamentally expressed and novel concept of nanotechnology (and related nano-bio-science) *does exist* as a particular application of the universal science of complexity [36,63,64], but is apparently lost on the background of superficial, money-driven publicity and deceptive successes of blind and dangerous empiricism.

A similar loss occurs in a yet larger field of *biological applications* of "exact" sciences, where their usual, unitary doctrine cannot explain the specific life properties even qualitatively, but proposes instead an infinite number of over-simplified mechanistic imitations of living system dynamics and rejects a realistic analysis providing unreduced life properties as manifestations of high enough levels of universal, interaction-driven dynamic complexity [35,36,64,65].

The same "sale" of nonexistent and improbable science results at a super-high price dominates in the field of controlled *thermo-nuclear fusion* for energy production that suffered from serious difficulties in a previous period (the end of the last century), but now has won a new, huge support (ITER project), despite the absence of practical progress or even theoretical solution. And here again, the real, underlying problem is due to irreducible dynamic complexity effects that just cannot be properly treated within the unitary science doctrine in principle. Not only strong and diverse plasma instabilities (due to the *genuine*, rather than simulative chaos) create particular difficulties in development of intrinsically inefficient hot fusion schemes, but much more efficient and promising approach of *cold fusion* can be formulated exclusively in terms of complex behaviour and therefore, not surprisingly, is either totally neglected, or pushed to a far margin of official science activity (on the background of multi-billion support for provably inefficient hot fusion). Needless to recall, we deal here with not only practically appealing, but urgently needed application of global importance; and still the official science machine prefers to support its "best" (i.e. self-selected) *people* interests, rather than the objective *science* quest and related interests of humanity.

A.P. KirilyukAnd finally, as if in order to definitely kill any remaining hope for occasional knowledge progress in the epoch of the end of science [3], the official science establishment gives a very strong support to a major "new science" imitation in the form of *computer (simulation) science* (see e.g. [88,89]) and its extremely vast scope of applications ("everything can be put in a computer" and simulated). Even apart from the evident fact that a "computer experiment" cannot provide in itself any additional understanding (while it is far less precise than real observation results and often simply unrealistic), the unreduced, multivalued dynamics analysis reveals the fundamental deficiency of such basically single-valued imitations (cf. section 2.1.1) prone to multiple instabilities and related arbitrary large deviations from real phenomena. A characteristic example of such glaring inefficiency of "computer science" is provided by various simulations of the "system Earth" behaviour in relation to quickly growing ecological problems (e.g. [90]): after practically unlimited financial investments into the field one gets only the result that could be clearly expected from the beginning: the predicted "effect" is of the same order as the differences between various "supercomputer" simulation results, so that in the end one still can rely exclusively upon real system observations.

In all these cases, a logically strange but inherent property of the mechanistic science approach appears in its ultimately absurd form: the official positivism imitates everything it can using all practically accessible, ever more perfect *tools* of *purely empirical* technology, *irrespective of the obtained results utility* or any *real scientific* purpose of their production (now practically absent). It is the *empirical tool technology* that becomes the purpose in itself. One deals here with infinitely multiplying and cycling circles of "trial-and-error" efforts looking "promising", due to the growing *technical* possibilities of new tools, but in reality dropping dramatically in efficiency down to practical zero because of the "exponentially huge", practically infinite number of interaction possibilities within every "truly complex" (large enough) system dynamics [36,63-66]. The resulting deep impasse is evident: no progress is possible within the officially imposed science paradigm, and the more is the power (and cost!) of technical tools applied, the smaller is the hope to get out of vicious circles of unitary thinking. The epoch of blind empiricism is finished and it becomes really dangerous now, but still persists without practically visible limits, selfishly suppressing any attempt of provably efficient knowledge development. Only decisive, qualitatively big transition to the unreduced analysis of real, multivalued system dynamics can put an end to exponentially growing expenditures for successively failing, practically fraudulent giga-projects and open the urgently needed era of causally complete solutions to "difficult", and now critically stagnating, problems.

### 2.1.4. True lie, or post-modern science: unlimited complexity imitations

As even a quick glance at the official science landscape reveals a hierarchy of terrible deficiencies and critical stagnation of results outlined above, the ending unitary science System tries to preserve its absolute dominance (despite everything!) by making "concessions" to the evidently missing complexity in the form of mathematically incorrect, ultimately loose and speculative "science of complexity" often presented as the necessary "new paradigm". Since in fact there is nothing fundamentally, scientifically new in that pseudo-philosophical "paradigm" totally remaining within the same, severely limited space of dynamically single-valued approach, the final result tends to the natural limiting point of the bankrupt science doctrine, the so-called "post-modern science", made of practically arbitrary, senseless plays on pseudo-scientific words and symbols that not only has nothing to do with real system dynamics, but does not even pretend for it anymore, allowing for any evidently incorrect play with mathematical symbols and rules and considering purely verbal promises of a "new life" as the *only* desirable and possible result of scientific activity.

We shall not repeat here the detailed description of unitary imitations of complexity, their evident deficiency and catastrophic result of their artificial domination in practically all official science institutions. One can find those details in refs. [35-37] (accompanied by further elaboration in [57-66]) and a correct description of the resulting "perplexity" in popular accounts [3,4]. It is enough to mention the underlying basic deficiency of unitary science analysis replacing the huge plurality of permanently changing system realisations by only one, "averaged" (or even arbitrary) realisation representing the single possible, unchangeable system version (see section 2.1.1). It means that *any* unitary science structure, *including* those from its *complexity imitations*, has *strictly zero* value of *unreduced*, universally defined dynamic complexity, one of the consequences being complete perplexity and disorder with multiple competing complexity definitions of unitary "science of complexity" (and no one of them has the necessary, universal properties). Of course, that omnipresent deficiency becomes much more evident in the analysis of "truly complex", multi-component and characteristically "soft", unpredictably changing systems (culminating in living and intelligent ones), but it is interesting that even such striking contradiction between reality and its "scientific" image does not stop unitary imitations of positivistic doctrine that does not see any problem in e.g. a living (and thinking) organism imitations by a sequence of smooth, fixed and totally abstract mathematical structures.

As a particular example, one can evoke the unitary science notion of "chaos without chaos" based on evidently incorrect extension of a local series expansion ("Lyapunov exponents") leading to completely erroneous and conceptually flawed idea about real system evolution, instability and (false) "chaoticity", shifting the origin of randomness to "poorly known" initial conditions. That "exponentially amplified" uncertainty would give rise to only false, rather than real randomness that would, in addition, depend on time, thus making system chaoticity a (strongly) time-dependent issue. In the case of *quantum* (Hamiltonian) system dynamics, that contradiction attains its paroxysm in the form of *inevitably* regular behaviour of expected chaotic systems, in (false) conflict with the correspondence principle, while the unreduced, dynamically multivalued interaction analysis provides explicit and consistent problem solution in the form of universal origin of genuine, time-independent dynamic randomness in quantum and classical systems (see [35,36,59] and references therein). Other cases of official chaos without randomness include various "chaotic" or "strange" attractors that always describe a *single* possible system trajectory in an *abstract* (phase) space that can only approach its eventual "equilibrium" state, without any true randomness. Genuine chaoticity origin is replaced here by incorrect Lyapunov exponent use accompanied by "intuitively" estimated computer simulation results that being a computer analogue of natural experiments may reproduce, in principle, an unpredictable part of real system chaoticity but without revealing its true origin and characteristics. After which such "well-established" chaoticity doctrine is applied to a huge variety of systems, from few-body interactions to social dynamics, intelligence, and ecological systems.

And because such arbitrary deviations from elementary consistency are not only possible but dominating in the official science practice, it becomes also possible and quite "natural" to pass to *explicitly* and *officially* unlimited deviations, in the form of pure play on words of open post-modern science that doesn't need any more even those false justifications and derivations of "rigorous" unitary science and can make arbitrary guesses based on borrowed "tricky" terminology, without any understanding (let alone direct verification) of its meaning and origin (see e.g. [19-21]). Those post-modern games of ultimate knowledge destruction often hide themselves behind a superficial demand of "interdisciplinarity" (e.g. [56]) based on a "felt" necessity to transgress traditional disciplinary boundaries but actually reduced to *unlimited inconsistency* and arbitrary application of any "model" or notion to any phenomenon.‡‡ As the intrinsically complete interaction analysis of the universal science of complexity clearly shows, conventional disciplinary ruptures directly originate in the severely limited picture of *zero-dimensional reality projection* of the unitary science, while the full, *dynamically multivalued* vision of reality has no such problem in principle and provides "coherent", logically correct transitions between "disciplinary" views extended now to *levels of complexity* of the unreduced nature dynamics as confirmed by very diverse applications [35-37,57-66]. However, as official science sticks to the above over-simplified imitations of complexity, one obtains as a result not the promised unified, interdisciplinary knowledge (and *harmonious* society based on *such* knowledge), but rather knowledge and society based on the generally accepted, publicly supported lie and fraudulent imitations without limit.

---

‡‡ Thus, any limited mathematical tool or model can boldly and directly be applied to description of any higher-complexity phenomenon (such as life, intelligence, consciousness), as if its authors ignore the evident and essential difference between the tool simplicity and complexity of real structure it is applied to (see e.g. [91,92] for only one recent series of the kind). Nevertheless, it is those explicit and terribly incorrect *imitations* of complexity (often obtained by direct, and always silent, simplification of its unreduced description, e.g. [35,36,60,65]) that obtain generous support of the official science establishment.

## 2.2. Doctrinaire Science Organisation: The Curse of Unitary Paradigm

As follows from previous sections 2.1.1-4, the official science system supports exclusively the evidently inefficient approach of the unitary (dynamically single-valued), or "mechanistic", paradigm, despite quite explicit and quickly growing dangers from such science practice for the scientific enterprise itself and the related technical, economical, and social development on the whole. Because of blind, but technically ultimately high and therefore destructive power of unitary science, hitting today the *whole scale* of natural system complexity, the very survival (let alone sustainable progress) of human species after such "experimentation" becomes a more and more questionable issue (section 2.1.2) clearly felt as such by official science leaders [39]. If all of it is so clear, then why does the self-destructive enterprise continue so prosperously its activity profiting from a US$ trillion-scale, always growing yearly support? The deeply rooted, intrinsic limitations of *unitary thinking* trying to mechanistically simplify everything it touches upon is one general reason for it.

A related, more practically involved reason for the persisting domination of the suicidal doctrine is its *organisation* in a *Unitary System* that naturally reproduces in its centralised hierarchy the glaring defects of unitary knowledge content: mechanistic simplification, rigidity, ruptures and totalitarian stagnation of development (in other words, such science, such organisation of science). That *effectively* centralised, *administrative* kind of human activity organisation has proven its total inefficiency in general social life organisation by the recent spectacular fall of the command economic and social system in the Communist Block of countries (Soviet Union and its satellites). But exactly the same kind of system still strangely dominates in organisation of science in any country, where it is realised by the world *intellectual elite* than should at least understand the evident fact, if not the origin, of the administrative system failure. By analogy with the last phase of Communist regime existence, such strange "inertia" reveals the ultimately deep level of *corruption* within the Unitary System of science organisation, where superior hierarchy layers are paid enough (actually by themselves!) just in order to preserve the "nourishing" System, irrespective of its production efficiency (which is an example of *negative* "self-organisation" reduced here to dirty *privatization of truth*). Various versions of strong, dominating corruption at all levels of modern science, varying from open "intellectual" slavery to explicit or implicit scientific fraud and plagiarism, are extensively described in various sources and publications [1-38],[§§] and still there is no visible change of tendency or at least attempt of science organisation change, which demonstrates again the huge negative power of that *practically implemented* mode of unitary, positivistic thinking as well as the necessity of a *unified, revolutionary* kind of change in *both* scientific knowledge content and organisation.

It is rather evident that the only reasonable kind of such revolutionary change is the unified transition from unitary to realistic science content (exemplified by the universal science of complexity) and from administrative to free-production, creative and individually structured science organisation [35-38] (see section 3 for more details). Just as modern, unitary science structure closely corresponds to its mechanistic content, the free-interaction kind of science organisation at the forthcoming higher level is directly related to the new science content (unreduced complexity of real system dynamics), so that the new science of complexity can be successfully applied to (consistent) understanding of science development itself (which can never be the case for unitary science content and development). Another feature confirming not only the necessity but also reality of deep science organisation change is the possibility to perform it starting from *small enough* volumes of a big enough change (as it happens in higher-order phase transitions in physical systems, contrary to simultaneous change in the whole system volume in the case of first-order transition).

Correspondingly, there is no sense to try to increase science efficiency by "ameliorating" its organisation details without changing the system as such (this kind of proposition inevitably dominates in presented official science ideas about its own reform). It is at this point that *rigorously substantiated* and *causally complete* results of the universal science of complexity on such "truly complex" structure evolution provide the *uniquely reliable* basis for practically important actions.

---

[§§] Beyond any public recognition, private communication with professional colleagues, which is especially efficient for *such* "contradictory" issue discussion, shows convincingly that published information demonstrates only "the tip of the iceberg" of science organisation and practice problems.

## 2.3. Science of Lie: The Ultimate Deadlock of Scholastic Knowledge

As a result of presented analysis of official, positivistic science paradigm (in both its content and organisation), sections 2.1-2, one can see that the clearly perceived modern change of scientific knowledge (e.g. [1-3,19-23,39-49]) should be specified as indeed the ultimately deep, qualitative and "final" crisis of the whole existing kind of fundamental science, i.e. the true end of *unitary* science, and the closely related transition to superior kind of knowledge (see also section 3 below).

Major features of usual, unitary science basis, viewed now from the new, causally complete knowledge perspective (section 2.1.1), reveal the fundamental, inevitable, and *rigorously specified* limitations of that *particular*, very special kind of knowledge in the form of its *effectively zero-dimensional* (dynamically single-valued) vision of all studied objects and phenomena, which is the real, concrete reason for the now exhausted possibilities of its further development and resulting impasse, or "end", of science. The necessity and origin of the new, causally complete and totally realistic kind of knowledge based on the *truly exact* analysis of *unreduced*, dynamically multivalued behaviour of real systems becomes equally clear, thus completing the emerging picture of objective relation between real world structure and its reflection in human knowledge.

The ultimately big scale and importance of the difference between usual (unitary and fixed) and new (realistic and creative) kinds of knowledge should not be either underestimated theoretically (as being due only to "knowledge refinement" or superficial "interdisciplinarity") or overestimated practically (as impossibility of further science development at a *superior* level, cf. [3]). The huge scale of modern science impasse and related transition to superior kind of knowledge is emphasized by the fact that today's change terminates scholar (and *scholastic*!) science development during its *whole history* of more than three centuries, starting from Newton's paradigm of a "technically adequate" description that does *not* need additional, complete *explanation* in principle ("hypotheses non fingo"). It is precisely that, very special science content, practice and attitude later summarised in the form of *positivistic science* doctrine that reaches now the ultimately deep, final impasse and contradiction to the necessity of genuine understanding of real, arbitrary complex system dynamics.

One can easily understand the origin of major observed properties of thus specified unitary science, its *scholastic dogmatism* and dramatic degradation from "plausible", technically correct imitations ("models") of reality to "scientifically packed" but obvious *lie*, i.e. *arbitrarily large* separation between "models" and real structures they are supposed to describe.

Dogmatism results from the dynamically single-valued, over-simplified "projection" of reality within unitary science paradigm that does not leave any place for a flexible interplay between multiple, often "opposite" system properties readily observed in nature (recall e.g. "wave-particle duality" observed already at the lowest levels of real world dynamics and remaining mysteriously inexplicable in the allegedly "rigorous" approach of official science). By contrast, such naturally occurring interplay constitutes a major, intrinsic property of unreduced, dynamically multivalued system behaviour in the universal science of complexity [35-37,57-66]. It also gives rise to *dynamic complexity development* from its lower to higher levels that provides a universal and causally complete solution to the problem of "interdisciplinary" vision and "reduction" of higher-level properties to lower-level interactions (or dynamically *emergent* system properties). There can be no more place here for a dogmatic fixation on a particular system property, quality, or feature.

In a similar way, the effectively zero-dimensional, point-like reality projection of unitary science cannot avoid *arbitrary large deformations* with respect to the unreduced, "multi-dimensional" (dynamically multivalued) structure of reality, where such deformations naturally grow from simpler (more fundamental) to higher-complexity systems (which explains an otherwise "strange" combination of a *relative* success of unitary paradigm application to simple physical systems and its catastrophically growing failure to provide an equally "exact" description of higher-complexity systems). Such surrealistically big deformations of reality, multiplied by their exponentially growing number and totally abstract origin create today the situation of a "science of lie", i.e. a form of allegedly "objective" knowledge but where the criteria and very notion of truth become "infinitely" smeared ("everything is possible") [3,4,19-21,23,24]. But as that "post-modern" state of "life after death" in science is accompanied by a particularly prosperous development of *empirically based* technology, one obtains a yet more surrealistic situation where that ultimately

A.P. Kirilyuk

wrong and practically dangerous science of lie is taken as the uniquely suitable basis for the expected "knowledge-based society". One should obtain thus a very "futuristic" society based on lie! (It may already be the case, looking at modern tendencies in "developed" countries...)

The transition from apparently prosperous, though maybe "non-ideal" fundamental science to its modern version of ultimately uncertain "fairy tales" with no relation to reality had *silently* happened somewhere in the middle of the 20$^{th}$ century, while the official science *status* has remained at its highest peak till recently by simple "inertia" effects of habitual propaganda of "evident" advantages of knowledge and its progress. In the meanwhile, another, *purely empirical* development of *technological tools* of science has attained equally critical (and also hardly recognised) point just in the same epoch, where they could, for the first time in human history, directly touch and modify the *deepest levels* of natural system complexity, from elementary particles to biological and ecological (as well as social) systems. In that way, a silent, but catastrophic drop of understanding of real system structure and dynamics coincided with the equally dramatic (and unrecognised) increase of the purely empirical power to change them *completely*. It does not seem difficult to see that the *cumulative effect* of those two major changes acting in the *same direction* leads to a *major transition*, in the middle of 20$^{th}$ century, from generally, externally useful (and therefore also relatively consistent, prosperous, creative) science to its ultimately dangerous, practically dead, and critically decaying version today, even though one deals in both cases with exactly the same, positivistic, unitary, mechanistic science doctrine. Clear recognition of that crucially important transition at least now, when the risk to suffer from its negative consequences is at its maximum, is a necessary condition for any further knowledge and civilisation development that can only take the form of revolutionary transition to a superior kind of knowledge (section 3).

As pointed out above (section 2.2), the observed sharp crisis in official science practice and organisation is closely related to the problem of its unitary content, so that both scientific knowledge content and organisation can break the current deadlock and start up a new, unlimited progress stage by the unified transition to the unreduced vision and analysis of real world phenomena as exemplified by the already realised framework of the universal science of complexity [35-38,57-66].

## 3. From Dogmatic to Creative Knowledge: Revolution of Complexity

### 3.1. Science of Truth: Intrinsically Consistent Knowledge

It is not difficult to see why further science development is impossible without a crucial progress towards genuine understanding of real interaction processes: the latter dominate the whole scope of modern practical applications where both intensity and "interconnectedness" of various interactions grow now without visible limits (it is a widely understood *globalisation* phenomenon). But as it is universally demonstrated in the unreduced complexity analysis [35-37,57,58], such real, arbitrarily large and intense interaction processes always give rise to rigorously specified dynamic complexity, in the form of permanent realisation change in a truly chaotic order. It becomes evident, therefore, that the highly needed and even *empirically*, tentatively emerging knowledge revolution mentioned above can only involve transition from the now dominating unitary, dynamically single-valued (zero-complexity) science approach and thinking to the unreduced, dynamically multivalued, or complex, behaviour of real systems and its qualitatively extended analysis. The emerging summary of ever growing number of "unsolvable" fundamental problems (sections 2.1.1-4) just directly demonstrates those "missing dimensions" of the official science framework. In other words, the forthcoming knowledge revolution is clearly specified now as the *revolution of complexity* understood as a *practical transition* to *superior complexity level* not only in formal, scholar knowledge, but also in various applications and general attitudes. Needless to say, such complexity revolution is very far from its bubbling unitary imitations that cannot provide even a clear definition of the main quantity of interest, complexity itself [3,4], let alone solve accumulating real problems. Talking very loudly about "the century of complexity", unitary science priests, provided with maximum material support in countless "advanced" institutions, in practice only submerge ever deeper in over-simplified, totally abstract games with mathematical structures of *zero complexity* (see sections 2.1.1-4), which clearly demonstrates their real intentions and capacities.

Despite that purely subjective suppression of the unreduced complexity by the whole official science establishment, its triumphant advent is inevitable at the progressive development branch, with the only possible alternative of destructive "death branch" [37]. Clear signs of the latter are already visible in the content and practice of official science (sections 2.1-3) just because it resists obstinately to the progressive tendency. In the meanwhile, one can objectively estimate, already now, the main properties of the forthcoming new knowledge and preceding complexity revolution.

Just as severely limited projections of unitary science lead inevitably to what can be called, due to essential and irreducible separation from reality, science of lie (section 2.3), the intrinsically complete knowledge of the unreduced science of complexity leads to the *science of truth* containing no ambiguous, abstract "models" and criteria of their validity varying subjectively in proportion to a personal "push" of one or another "eminent" scientist. The absolute, *objective criterion of truth* of that new, unreduced form of knowledge is its *total consistency*, i.e. absence of any noticeable contradictions in the *entire system of correlations* within the available (in principle, growing) volume of knowledge [35]. It is very different from the situation with irreducibly separated "models" of unitary science where each model (a point-like projection) can reproduce some system properties but fails with other, equally important ones (e.g. wave-particle duality, or in general complementarity, in particle physics). Another, quantitative expression of the same criterion of truth states simply that unreduced complexity of "mental" constructions of knowledge (i.e. our "understanding" of reality) should be *equal* to that of real world objects (including brain-reality connections as they represent but a particular case of complex-dynamic interaction). In other words, science of truth provides, in agreement with its name, the *truly exact* version of the world structure and dynamics (within a growing volume of empirical data), being thus the *intrinsically complete* (totally realistic and consistent) form of knowledge. By contrast, as follows from the same criterion, unitary science knowledge represents the *largest possible* deviation from truth as it has the minimum possible, zero value of unreduced complexity (while real structure complexity is always high, even for the simplest structures [35,60-62]). This is the real reason for a strange combination of visible, mainly empirical "successes" of positivistic science and its *growing* failure to *explain* even the simplest, most fundamental observed entities and properties (those of time and space, elementary particles, etc.).

Because of that special nature of new knowledge and its difference from the previous level of unitary science, the transition to the unreduced complexity science also has quite special properties of the *last scientific revolution* [37]. According to the well-known idea of scientific revolution, or "paradigm change", by Thomas Kuhn [53], the process of knowledge acquisition has a quite uneven and very "contradictory" structure involving antagonistic fight between emerging new ideas and inertia of currently "established" knowledge. Although the nonuniform character of any progress is confirmed by the causally complete analysis of the universal science of complexity [37], it becomes also evident that unitary science "progress" has been occurring mainly in the form of *empirical discoveries* due to *technical* progress of research *instruments*, while the essential *understanding* of reality has always remained at the same superficial, lowest possible, zero-complexity level. This is exactly why the change to a new paradigm within unitary science has that "antagonistic" character: without genuine understanding of reality, each new discovery appears as a "miracle" difficult to believe in at the beginning, until everybody can empirically verify its reality (recall microbial disease origin discovery by Pasteur, difficult emergence of the "new physics", etc.). But the discovered "new" reality is *not* consistently explained either, usually it would even be *less* understandable than previously known "simpler" objects, which endows unitary science "progress" with a "contradictory" property of *growing accumulation of mysteries and unsolved problems*, while even old, previous-level problems do *not* really obtain their clarification after the advent of a new paradigm (postulated Newton's laws were not provided with any deeper understanding after the appearance of the "new physics" that has instead brought about its own mysteries growing in number and scale until now!).

One can see now why the forthcoming transition from unitary science to the unreduced science of complexity, or revolution of complexity, has quite a different character with respect to Kuhnian scientific revolutions and gives rise to a *qualitatively new kind of knowledge progress*, in the form of *permanent extension* of previous *understanding* of reality (*growing consistency* rather than mystification) without those painful ruptures and antagonistic contradictions between old and new knowledge. It is the expression of the *intrinsically creative* nature of the *causally complete* knowledge of unreduced complexity science. After transition to that superior kind of knowledge, Kuhnian scientific revolutions become impossible as such: uneven knowledge progress proceeds without ruptures and inexplicable "miracles". That's why the revolution of complexity is the *last scientific revolution* or, more precisely, it puts an end to unitary scientific revolutions and opens a qualitatively *new way* of *intrinsically sustainable development* of a *new kind* of knowledge.

It is interesting that the basis for that new kind of knowledge has actually been created by René Descartes yet before the appearance of Newtonian, positivistic science that "does not invent hypotheses" (i.e. explanations). Indeed, it is easy to see [35] that Cartesian science (e.g. [93,94]) was just very strongly oriented to production of the *most complete explanation* of observed phenomena (which even sometimes led Descartes to erroneous theories, in the absence of necessary experimental data). Related Cartesian approach *creativity* (due to his famous concept of constructive *doubt* [93]) and *unified* ("interdisciplinary") character of resulting knowledge only confirm similarity with the unreduced complexity science and the fact that such another kind of knowledge was initiated before Newtonian positivism, at the very beginning of modern science as such. Newton later argued in favour of his apparently exact (and therefore visibly efficient) *quantitative*, mathematical analysis, which he absolutised to the idea that one did *not* need *anything* else than "exact" mathematical models, any additional explanations being but unnecessary "interpretations". Whereas Newtonian science application to (major) planetary orbit description may seem to justify such attitude, its failure to solve already a three-body problem, let alone provide an adequate description of "truly complex", e.g. biological, systems shows convincingly that "everything is not so simple" in the world containing a lot more than combination of regular Platonic figures of the unitary paradigm (cf. [74,75]). The problem was, therefore, not so much in that Newton had strongly imposed his indeed apparently "perfect" (at that time!) theory, but that in was not later properly completed and extended again to the lost consistency of initial, Cartesian science paradigm, not even at the moment of *explicit complexity emergence* in the "new physics" discoveries at the beginning of the 20$^{th}$ century (chaotic dynamics, stochasticity, quantum and relativistic behaviour, cf. [35,60]). It is only today that one can see the true meaning of the lost "method" of René Descartes, and even now, almost four centuries after his work, the huge machine of official science, provided with "miraculous" technical power, continues to operate senselessly for the evidently fruitless mechanistic doctrine unable to ascend to the level of intrinsically complete knowledge founded by Descartes at the dawn of scientific age.***

That "contradictory" way of real knowledge development shows that it is far from being finished and unreduced, realistic knowledge has not started yet its genuine, unrestricted progress, remaining under the pressure of *purely subjective prejudice* of a pre-scientific epoch of *scholastic*, religion-based kind of knowledge (superficial *interpretation* of a *basically fixed* doctrine based on a blind belief). New knowledge of the unreduced science of complexity, being modern realisation and well-specified renaissance of anti-scholastic, Cartesian tradition, is a *permanently developing*, open kind of knowledge devoid of artificial, subjective limitations (such as the "necessary" mystification of modern "rigorous" science) and guided only by the universal criterion of truth, the *totally consistent understanding* of reality within the whole accessible, ever growing volume of knowledge.

### 3.2. New Science Organisation: Interaction-Driven Creation

It is evident that the forthcoming superior kind of knowledge needs a new organisation, which should be as different from today's official science establishment (section 2.2) as the new science content (and real world dynamics) is different from the unitary knowledge projection. The intrinsically creative organisation of new knowledge can only be realised in the form of a free-interaction system of quasi-independent units of knowledge creation, support and propagation [38]. Those *independent science enterprises* would mainly be based on *individual* researcher efforts and small teams, but being freely composed and recomposed structures, will certainly include various hierarchical unifications (e.g. within particular projects), as well as accompanying *knowledge management enterprises* that will make their business on optimal financial support for *knowledge creation enterprises* and their result application, dissemination, and public estimation.

---

*** Even worse, the bankrupt unitary science tends to attribute to Descartes a particular attachment to its own, mechanistic approach, while reserving to itself the honour of "extension" of such simplified vision! It also tends to see Descartes as a "mere philosopher", rather than scientist, despite the well-known fact that he practically elaborated the mathematical basis of science itself, as well as much of "Newtonian" motion laws and various other results confirmed later, while being strongly restricted, at his time, by poor experimental possibilities. That is another instructive lesson of knowledge development showing that unitary science of lie unfortunately confirms, once again, its major character and purpose. It is interesting that Descartes prodigiously foresaw even that sad destiny of his approach when he wrote that many centuries should pass before science would be able to properly develop the principles of his approach.

That transition from today's unitary, rigidly centralised and administrative science organisation to the superior system of freely, constructively interacting knowledge creation units is analogous to the transition from a command economy within a totalitarian political system to a (regulated) free-market kind of economical life, with similar advantages in creative power and efficiency. There will be no more formal, unnecessary, subjectively driven limitations for either scientific creation itself or science organisation development that now becomes an *integral part* of creative knowledge production [38] instead of modern inefficient and corrupt bureaucracy serving just to *limit* knowledge development in favour of selfish interests of reigning feudal priests of science and their perverted clans. The acute crisis of modern science [1-49] widely recognised even within its unitary "establishment" shows convincingly that this kind of transition in knowledge organisation is more than urgent, if any knowledge progress is to be preserved at all. Only free-interaction kind of development can provide a truly efficient science organisation that becomes vitally necessary today for the whole civilisation progress, especially taking into account the huge and ever growing role of powerful modern technologies that easily become frustrated and dangerous without a guiding contribution from a constructive, problem-solving and *independent* (i.e. truly free) fundamental research (section 2.3). That is the only possible solution to a stagnating and much discussed problem of the *freedom of scientific research* (see e.g. [51,52]), which cannot be solved by any "reform" of traditional, unitary organisation of science. Indeed, contrary to naturally established practices of a free-market system, inefficient administrative "research" without any real progress can (and does!) continue for many decades without even any attempt to change its leaders and content.

Similar to higher-order phase transitions, that "symmetry change" in science organisation can occur gradually by its volume, but remaining qualitatively strong there where it does happen. A qualitatively big transition becomes thus more practically feasible than a simultaneous transition in the whole volume (first-kind phase transitions). Starting there where it is more probable and necessary, the transition to "distributed" and intrinsically creative science organisation will grow by its natural success and include interaction patterns and enterprises of various suitable scale and function. As there is no reasonable limit to such kind of intrinsically creative development of science content and organisation, one can be sure about its sustainability that doesn't need any additional, centralised kind of administration, as opposed to any version of unitary organisation. Note that first attempts to create interactive, quasi-independent units of research of a new, "liberal" kind already appear and grow everywhere, even though they still rely essentially upon usual science organisation.

Finally, it is worthy of mentioning that such transition to a superior, qualitatively more efficient kind of science organisation can serve as a prototype of a similar, equally necessary transition in the whole social structure suffering today from a deep development crisis [37].

## 4. Sustainable Development Based on Causally Complete Knowledge

While the necessity of transition to another, qualitatively different kind of civilisation development becomes evident from various perspectives and within different approaches giving rise, in particular, to the ecologically motivated *sustainable development* idea [90], it seems yet to be poorly recognised that such important change can only be based on the equally deep progress of underlying knowledge, so that the desired truly sustainable civilisation development can be *uniquely* realised in the form of society based essentially on a *new kind of science* ensuring causally complete, totally *consistent understanding* of *all practically modified systems* [37]. As the already realised applications of universal science of complexity convincingly demonstrate [35-37,57-66], that kind of knowledge also uniquely provides the causally complete understanding of fundamental universe structure, laws and purpose, from elementary particles and cosmological problems to life, intelligence, consciousness and their development. It is such unreduced, reality-based vision of the forthcoming change of science that should guide its modern development, as opposed to obscure manipulations of unitary science scribes remaining totally closed within their self-interested, narrow doctrines and abstract models separated from real life and related purposes of human progress.

The best science advances have always been driven by intrinsic, individual creativity and constructive interaction within the whole civilisation development. But those could only be rare, "enlightenment" moments in the dominating kingdom of scholastic unitary thinking. And in today's epoch of "material life" triumph, fundamental knowledge as such has lost its creative character, superior purposes and has become just an imitative, parasitic and unpopular appendage to flourishing empirical technologies. There is no positive solution on that way of quickly advancing decadence, for either science, or civilisation whose development it should guide. Any hope for usual, evolutionary progress by small steps within the existing system is vain, that is the definite conclusion of both rigorous analysis of the universal science of complexity (applied now to the system of science or civilisation as a whole) and accompanying qualitative considerations. The last scientific revolution outlined in this paper is a unified and uniquely consistent change to another, progressive branch of development of knowledge and civilisation based on the power of that qualitatively new knowledge.

The Last Scientific Revolution